\documentclass[twocolumn,prd,groupedaddress,nofootinbib,showpacs]
{revtex4}

\usepackage{latexsym}
\usepackage{amssymb}
\usepackage{graphicx}
\usepackage{dcolumn}
\usepackage{bm}

\begin{document}

\title{Statistical transmutation of quantum bosonic strings coupled to
general four-dimensional Chern-Simons theory}

\author{J. Barcelos-Neto}
\email{barcelos@if.ufrj.br}
\author{E.C. Marino}
\email{marino@if.ufrj.br}
\affiliation{Instituto de F\'{\i}sica\\
Universidade Federal do Rio de Janeiro, RJ 21945-970 -- Brazil}

\date{\today}

\begin{abstract}

A bosonic string coupled to the generalized Chern-Simons theory in
3+1D acquires a magnetic field along itself, when it is closed, and a
topological charge at its extremity, when it is open. We construct the
creation operators for the full quantum field states associated to
these strings and determine the dual algebra satisfied by them. We
show that the creation operator fo the composite state of a quantum
closed bosonic string, bearing a magnetic flux, and a topologically
charged open bosonic string, possesses generalized statistics. The
relation of our results with previous approaches to the problem is
also established.

\end{abstract}

\pacs{11.10.Ef, 11.15.-q, 11.25.-w, 05.90.+m}

\maketitle

\section{Introduction}

The obtainment of a full quantum field theory of strings would require
the construction of the complete creation operator for the string and
the evaluation of its correlation functions. Nevertheless, a full
quantum description of the statistical transmutation undergone by
bosonic strings when coupled to certain fields, surprisingly, can be
achieved at an intermediate stage, where the fields attached to the
string by the interaction are fully quantized. The achievement of this
description, which may be considered as a partial field quantization
of the bosonic string is the main purpose of this work.

\medskip
Statistical transmutation of bosonic strings has been recently
investigated in various different frameworks \cite{fgt,gs,fg,m1}. It
has been shown in particular \cite{gs} that a closed spatial string
bearing a magnetic flux and moving in the presence of the field of a
classical topological charge would undergo statistical transmutation.
Here we provide the full quantum description of this fact in the
framework of the generalized Chern-Simons (CS) theory in 3+1
dimensions. This theory has been shown to belong to a hierarchy of
theories, related by dimensional reduction, which contains the usual
CS Lagrangian in 2+1D as a particular case \cite{cs3}. We start from
the observation that a bosonic string coupled to this theory bears a
magnetic flux along itself, whenever it is closed and a topological
charge at its extremity, whenever it is open. We then explicitly
construct the operators creating the full quantum field states, which
are attached to the string through the coupling. The creation operator
for the magnetic field along the string is basically the same one
considered previously in the framework of the BF-theory \cite{m1}. The
topological charge quantum field creation operator, however, is
specific for the CS-theory. We show that these operators indeed create
eigenstates of the corresponding appropriate operators with the
correct eigenfield configurations. We then show that the magnetic
field and topological charge creation operators satisfy a dual
algebra, which implies that their product possesses generalized
statistics. We also show that our result is the operator version of
the statistical transmutation of bosonic Nambu-Goto strings coupled to
CS theory, observed in \cite{fgt} within the functional integral
framework.

\medskip
We mention that there is another CS term at $D=3+1$ \cite{cs3}
involving a scalar and rank three gauge fields. However, this term is
not related to statistical transmutation because there is no
charge-flux system associated to it.

\medskip
The procedure we have presented in this paper can be naturally
extended to higher spacetime dimensions, where statistical
transmutation of membranes coupled to gauge fields of rank higher than
two should occur \cite{mb}.

\section{Chern-Simons Theory in 3+1D and Its Coupling to the String}
\renewcommand{\theequation}{2.\arabic{equation}}
\setcounter{equation}{0}

Chern-Simons theory in 3+1D involves a gauge vector field $A_\mu$ as
well as a gauge tensor field $B_{\mu\nu}$ (Kalb-Ramond) and is
characterized by the Lagrangian \cite{cs3,cs4}

\begin{equation}
{\cal L}_{CS}=\frac{1}{2}\,\epsilon^{\mu\nu\alpha\beta}A_{\mu }\,
\partial_{\nu}B_{\alpha\beta}.
\label{1}
\end{equation}

\noindent
Consider now a point particle, described by the current $j^\mu$ and a
string placed at the curve $C$ and associated to the current density

\begin{equation}
J^{\mu\nu}=\int_{S(C)} d^2\xi^{\mu\nu}\delta(x-\xi),
\label{2}
\end{equation}

\noindent
where ${S(C)}$ is the universe-sheet of the string. We may couple
these to the Chern-Simons Lagrangian in the usual way, namely,

\begin{equation}
{\cal L}=\frac{1}{2}\,
\epsilon^{\mu\nu\alpha\beta}A_{\mu }\partial_{\nu}B_{\alpha\beta}
-\frac{1}{2}\,J^{\mu\nu}B_{\mu\nu}-j^\mu A_\mu .
\label{3}
\end{equation}

The field equations associated to (\ref{3}) are

\begin{eqnarray}
&&J^{\mu\nu}=\epsilon^{\mu\nu\alpha\beta}\partial_\alpha A_\beta
\nonumber\\
&&j^\mu=\epsilon^{\mu\nu\alpha\beta}\partial_{\nu}B_{\alpha\beta}.
\label{4}
\end{eqnarray}

\noindent
We see that the string density along direction $i$ is given by

\begin{equation}
J^{0i}= \epsilon^{ijk}\partial_j A_k={\cal B}^i ,
\label{5}
\end{equation}

\noindent
implying the existence of a magnetic field $\vec{{\cal B}}$ along the
string. Note, however, that since magnetic field lines should be
closed, only closed strings can bear a magnetic flux. Also the second
field equation means that $j^\mu$ is identified with the topological
current, implying that a topological charge is associated to the
particle coupled to $A_{\mu }$. A very interesting case is the one in
which we take

\begin{equation}
j^\mu=\oint_C d\xi^\mu\delta^4(x - \xi) .
\label{5a}
\end{equation}

\noindent
where $C$ is the border of the universe-sheet of the string. For a
closed spatial string $j^0 = 0$, implying that this type of string
does not bear a topological charge, but only magnetic flux. For an
open spatial string, however, $j^0 \neq 0$ and we see that it carries
topological charge but no magnetic flux. Notice that, in this case,
$C$ is still closed in spite of the fact that the string is open.

\medskip
In what follows, we are going to study the operators creating the
quantum states corresponding to the above mentioned field
configurations. For this, we will need the basic commutation rules
for the theory described above, which are given by the expression

\begin{eqnarray}
&&[B^{ij}(\vec x,t),A^k(\vec y,t)]=i\,\epsilon^{ijk}
\delta(\vec x - \vec y),
\nonumber\\
&&[B^{ij},B^{kl}]=[A^{i},A^{j}]=0 .
\label{6}
\end{eqnarray}

\noindent
after elimination of the second-class constraints of the
theory (no gauge condition was used).

\section{The Magnetic Field and Topological Charge Operators}
\renewcommand{\theequation}{3.\arabic{equation}}
\setcounter{equation}{0}

Let us determine now the creation operators for the field
configurations corresponding to the closed string and point
topological charge. Firstly, we shall obtain the creation operator for
the magnetic field associated to a closed string at the spacelike
curve $C$. Following \cite{m1,m2}, we write

\begin{equation}
\sigma(C,t)=\exp\left\{\frac{ia}{2}\int_{S(C)}d^2\,
\xi^{ij}B^{ij}(\vec\xi,t)\right\}
\label{7}
\end{equation}

\noindent
where $d^2 \xi^{ij}$ ($i,j$ along $S(C)$) is the surface element of
$S(C)$, an arbitrary surface with border along $C$.

\medskip
According to the field equation (\ref{5}), the string is associated to
a magnetic flux along $C$. In order to characterize $\sigma(C)$ as a
quantum creation operator for the field configuration associated to
the string, let us evaluate $[{\cal B}^i,\sigma(C)]$. For this
purpose, we write $\sigma(C)\equiv e^{\alpha(C)}$ and, using
(\ref{6}), obtain

\begin{equation}
[{\cal B}^i(\vec x,t),\alpha(C)]=a\oint_C d\xi^i\,
\delta(\vec\xi-\vec x).
\label{9}
\end{equation}

\noindent
From this we get

\begin{equation}
[{\cal B}^i(\vec x,t),\sigma(C,t)]=\left[a\oint_C d\xi^{i}\,
\delta(\vec\xi-\vec x)\right]\,\sigma(C,t),
\label{10}
\end{equation}

\noindent
which implies

\begin{equation}
{\cal B}^i(\vec x)\vert\sigma(C)>=\left[a\oint_C d\xi^{i}\,
\delta(\vec\xi-\vec x)\right]\vert\sigma(C)> ,
\label{11}
\end{equation}

\noindent
which clearly shows that the operator $\sigma$ indeed creates an
eigenstate of the magnetic field with the correct
eigenfield-configuration.

\medskip
We now turn to the topological charge creation operator. Let us
consider

\begin{equation}
\mu(\vec x,t)=\exp\left\{-ib\int_{-\infty,L}^{\vec x}d\xi^{i}\,
A^{i}(\vec\xi,t)\right\},
\label{12}
\end{equation}

\noindent
where $L$ is an arbitrary line going from $-\infty$ to $\vec x$. We
are going to evaluate the commutator $[Q,\mu]$, where

\begin{equation}
Q=\int d^3x\,\epsilon^{ijk}\partial_{i}B_{jk}
\label{13}
\end{equation}

\noindent
is the topological charge operator. Writing
$\mu(\vec x,t)\equiv e^\beta(\vec x,t)$ and using (\ref{6}), we obtain

\begin{equation}
[Q,\beta(\vec y,t)]=b\int d^3x\,\delta(\vec x - \vec y),
\label{14}
\end{equation}

\noindent
which implies $[Q,\mu]=b\mu$ or $Q\vert\mu>=b\vert\mu>$, thus showing
that the operator $\mu$ carries $b$ units of topological charge.

\medskip
According to our considerations after Eq. (\ref{5a}), observe that we
may consider (\ref{12}) as the creation operator for the field
configuration associated to an open string at the curve $L$, which
bears a topological charge at its extremity $\vec x$.

\section{Generalized Statistics of Composite Strings with Topological
Charge and Magnetic Field}
\renewcommand{\theequation}{4.\arabic{equation}}
\setcounter{equation}{0}

Before studying the commutation rules of the composite topologically
charged string operators, let us determine the dual algebra satisfied
by the magnetic field and topological charge creation operators,
$\sigma \equiv e^\alpha$ and $\mu \equiv e^\beta$, given respectively
by (\ref{7}) and (\ref{12}). Using (\ref{6}), we obtain

\begin{equation}
[\alpha(C,t),\beta(\vec y,t)]=iab\int_{S(C)} d^2\xi^i
\int_{-\infty,L}^{\vec y} d\eta^i\,\delta(\vec\xi-\vec\eta),
\label{15}
\end{equation}

\noindent
where $d^2\xi^i\equiv\frac{1}{2}\epsilon^{ijk}d^2\xi^{ij}$.

\medskip
We now use the identity

\begin{equation}
\partial_{(\xi)}^i\left[\frac{1}{4\pi\vert\vec\xi-\vec x\vert}\right]
=\left\{\begin{array}{cc}
\epsilon^{ijk}\partial_{(\xi)}^j\varphi_k(\vec\xi-\vec x)
;&\vec\xi\not\in V_L\\
\int_{-\infty,L}^{\vec x}d\eta^i\delta(\vec\eta-\vec\xi)
;&\vec\xi\in V_L\\
\end{array}\right.
\label{16}
\end{equation}

\noindent
where $V_L$ is a cone of infinitesimal angle with vertex at $\vec \xi
= \vec x$ and axis along the line $L:(-\infty,\vec x)$ and $\vec
\varphi = \frac{1-\cos\theta}{r\sin\theta}\hat\varphi$, with $r=
\vert\vec\xi-\vec x\vert$. Because of the $delta$-function, observe
that (\ref{15}) is non vanishing only inside $V_L$, hence we can use
the second part of (\ref{16}) to write

\begin{eqnarray}
[\alpha(C,t),\beta(\vec y,t)]&=&-i\frac{ab}{4\pi}\int_{S(C)}
d^2\xi^i\,
\frac{(y-\xi)^i}{\vert\vec y-\vec\xi\vert^3}
\nonumber\\
&=&-i\frac{ab}{4\pi}\,\Omega(\vec y;C)
\label{17}
\end{eqnarray}

\noindent
where $\Omega(\vec y;C)$ is the solid angle comprised between $\vec y$
and the curve $C$. Since $[\alpha,\beta]$ is a c-number, it
immediately follows that

\begin{equation}
\sigma(C_x,t)\mu(\vec y,t)=\mu(\vec y,t)\sigma(C_x,t)
\exp\left\{-i\frac{ab}{4\pi}\Omega(\vec y;C)\right\}.
\label{18}
\end{equation}

\noindent
This is the dual algebra \cite{dual,odd} satisfied by the magnetic
string field and topological charge operators, which is identical to
the one satisfied by the corresponding operators in the B-F theory
\cite{m1}.

\medskip
We now consider the composite operator creating a
magnetic-field-bearing-topologically-charged string. Cho\-osing the
string as a circle $C_x$ centered at $\vec x$, this may be defined as

\begin{equation}
\psi(x;C_x;t)=\lim_{\vec x\rightarrow\vec y}
\mu(\vec x,t)\sigma(C_y,t)
\label{19}
\end{equation}

\noindent
Now, considering the property $\Omega(\vec x;C_y)-\Omega(\vec y;C_x)=
4\pi\epsilon\left(\Omega(\vec x;C_y)\right)$, where $\epsilon(x)$ is
the sign function, we get

\begin{eqnarray}
&&\psi(x;C_x;t)\psi(y;C_y;t)
\nonumber\\
&&\phantom{\psi(x;}
=e^{-iab\epsilon\left(\Omega(\vec x;C_y)\right)}
\psi(y;C_y;t)\psi(x;C_x;t).
\label{20}
\end{eqnarray}

\noindent
This commutation relation shows that the composite field operator
$\psi$, which creates the quantum field states with topological charge
and magnetic flux, associated to the composite of a closed and an open
string, possesses generalized statistics $S = \frac{ab}{2\pi}$ in the
four- dimensional Chern-Simons theory given by (\ref{3}). Suppose now
the quantum bosonic string, Nambu-Goto for instance, is created by a
certain bosonic operator $\Sigma$, in the absence of the CS coupling.
When the string is coupled to the CS Lagrangian, the appropriate
operator becomes $\Sigma\sigma$ or $\Sigma\mu$ (where $\sigma$ and
$\mu$ are given by (\ref{7}) and (\ref{12})), respectively, according
to whether the string is closed or open. Hence, the appropriate
operator for the composite closed-open string quantum states becomes
$\Sigma_C\sigma\Sigma_O\mu = \Sigma_C\Sigma_O\psi$. This has
generalized statistics $S$, according to (\ref{20}). The complete
description of the statistical transmutation of bosonic strings
coupled to CS theory in 3+1D, therefore, has been achieved at a full
quantum level.

\section{Remarks on the Functional Integral Formulation}

In a previous work \cite{fgt}, it has been shown that a bosonic string
coupled to four-dimensional Chern-Simons theory as in (\ref{3}), with
$j^\mu$ given by (\ref{5a}) undergoes Fermi-Bose transmutation. The
demonstration is made within the functional integral formulation,
where a sum is performed upon all possible universe-sheets $S(C)$
associated to $C$. We would like to remark that the result found in
\cite{fgt} is related to ours in an interesting way. In order to see
this, we note that the coupling with $A_{\mu}$ in (\ref{3}) involves a
sum over all values of the index $\mu$ and this fact is used in the
derivation of the fermionic algebra that leads to the Fermi-Bose
transmutation in \cite{fgt}. Observe now that the spacelike part of
the sums corresponds to a closed spatial string, which, as we saw,
bears a magnetic flux along it. The timelike part of the sums, on the
other hand, correspond to an open spatial string, which bears a point
topological charge at its tip, as we have seen. The whole sum in
(\ref{3}), therefore, comprises the composite state of a topologically
charged open string and a magnetic flux carrying closed string, which
is precisely the situation we have considered here from the operator
point of
view. Transmutation to a generalized statistics should be also
obtained in the framework used in \cite{fgt} by considering
arbitrary parameters in the couplings in (\ref{3}). We see that the
present work, therefore, unifies the results obtained in \cite{fgt}
and \cite{gs} and opens the possibility of evaluation of quantum
correlators for the operators describing the quantized field
associated to strings coupled to vector and tensor fields $A_{\mu}$
and $B_{\mu\nu}$.

\begin{acknowledgments}
This work was supported in part by CNPq, FAPERJ and
PRONEX-66.2002/1998-9
\end{acknowledgments}

\end{document}